\documentclass[conference]{IEEEtran}
\usepackage[nolist]{acronym}
\usepackage{graphics}
\usepackage{amsmath}

\usepackage{mathtools}
\usepackage{amsmath}
\usepackage{algorithm} 
\usepackage{algpseudocode}

\usepackage{tikz}
\usetikzlibrary{shapes,shadows,arrows}
\tikzstyle{decision} = [diamond, draw, fill= blue!50]
\tikzstyle{line} = [draw, -latex']
\tikzstyle{elli} = [draw, ellipse, fill=red!50, minimum height = 8mm]
\tikzstyle{block} = [draw, rectangle, fill= blue!50, text width=8em, text centered, minimum height = 15mm, node distance=5em]
\tikzstyle{line} = [draw,-latex']
\usepackage{amsmath}
\usepackage{tikz}
\usepackage{mathdots}
\usepackage{yhmath}
\usepackage{cancel}
\usepackage{color}
\usepackage{siunitx}
\usepackage{array}
\usepackage{multirow}
\usepackage{amssymb}
\usepackage{gensymb}
\usepackage{tabularx}
\usepackage{extarrows}
\usepackage{booktabs}
\usetikzlibrary{fadings}
\usetikzlibrary{patterns}
\usetikzlibrary{shadows.blur}
\usetikzlibrary{shapes}
\usepackage{subcaption}
\usepackage{amsmath, amssymb, bm}
\usepackage{caption}
\usepackage{subcaption}
\usepackage{amssymb}

\usepackage
{xcolor}
\usepackage{soul}
\begin{document}

\pagenumbering{arabic}

\title{Continuous Transfer Learning for UAV Communication-aware Trajectory Design}

\author{\IEEEauthorblockN{Chenrui Sun\IEEEauthorrefmark{1}, Gianluca Fontanesi\IEEEauthorrefmark{2}, Swarna Bindu Chetty\IEEEauthorrefmark{1}, Xuanyu Liang\IEEEauthorrefmark{1}, Berk Canberk\IEEEauthorrefmark{3} and Hamed Ahmadi\IEEEauthorrefmark{1}}
\\
\IEEEauthorrefmark{1}School of Physics Engineering and Technology, University of York, United Kingdom\\
\IEEEauthorrefmark{2}Interdisciplinary Centre for Security, Reliability, and Trust (SnT), Luxembourg\\
\IEEEauthorrefmark{3}, Edinbrough Napier University, Edinbrough, United Kingdom\\

}

\maketitle
\begin{abstract}

\ac{DRL} emerges as a prime solution for \ac{UAV} trajectory planning, offering proficiency in navigating high-dimensional spaces, adaptability to dynamic environments, and making sequential decisions based on real-time feedback. Despite these advantages, the use of \ac{DRL} for \ac{UAV} trajectory planning requires significant retraining when the UAV is confronted with a new environment, resulting in wasted resources and time. Therefore, it is essential to develop techniques that can reduce the overhead of retraining \ac{DRL} models, enabling them to adapt to constantly changing environments. This paper presents a novel method to reduce the need for extensive retraining using a \ac{DDQN} model as a pre-trained base, which is subsequently adapted to different urban environments through \ac{CTL}. Our method involves transferring the learned model weights and adapting the learning parameters, including the learning and exploration rates, to suit each new environment's specific characteristics. The effectiveness of our approach is validated in three scenarios, each with different levels of similarity. \ac{CTL} significantly improves learning speed and success rates compared to \ac{DDQN} models initiated from scratch. For similar environments, \ac{TL} improved stability, accelerated convergence by 65\%, and facilitated 35\% faster adaptation in dissimilar settings.

\end{abstract}
\IEEEpeerreviewmaketitle
\vspace{-0.2in}
\section{Introduction}
\acresetall
\acp{UAV} are increasingly employed in a broad spectrum of applications, including surveillance, agriculture, disaster management, and communication infrastructure maintenance~\cite{won2022survey}. 
Efficient planning and optimization of \ac{UAV} trajectories is critical, since the success of numerous \ac{UAV} applications, such as real-time data transmission and remote sensing, heavily relies on a reliable trajectory. One of the key challenges in \ac{UAV} trajectory is ensuring robust communication between \acp{UAV} and ground-based infrastructure, such as \acp{BS}.
This challenge becomes even more pronounced in complex urban environments with obstacles and interference.

Beside conventional optimization methods, \ac{RL} methods have emerged as suitable \ac{ML} solutions for \ac{UAV} trajectory planning due to its proficiency in handling high-dimensional spaces, adaptability to dynamic environments, and capability for sequential decision-making based on real-time interactions and feedback \cite{bithas2019survey}. 
The navigation of \acp{UAV}  from an initial location to a final destination through \ac{RL} has been extensively explored in various related works \cite{chenrui}.
\ac{RL} approaches for \ac{UAV} trajectory and communication scenarios proposed in literature vary from classical \ac{RL} \cite{yin2019intelligent}, to \ac{DRL} algorithms \cite{moon2021deep,zhu2021uav}.
Adding communication constraints, as discussed in \cite{ding20203d, 9625502}, these works address the challenge of frequency band allocation in \ac{UAV} trajectory design using \ac{DRL} to ensure equitable communication services. These works highlight the importance of frequency management in \ac{UAV} operations. To further improve performance of trajectory, in \cite{yang2019connectivity}, cellular-connected \ac{UAV} technology was utilized to enhance 3D communication coverage, employing \ac{DRL} with a model-based approach for trajectory optimization. 
The strength of \ac{DQL} lies in its ability to learn optimal policies for \ac{UAV} navigation through interaction with the environment, thus enabling precise trajectory planning and robust communication strategies. 

Nevertheless, the utilization of \ac{DQL} comes with limitations. Although it performs well in learning tasks within specific environments, these solutions are often tailored to particular locations or maps. 
Adapting these models to new tasks or different environments often requires extensive retraining, a process that can be both time-consuming and resource-intensive. This limitation is particularly evident in \ac{UAV} operations that demand rapid adaptation to new contexts. In fact, the challenges and dynamics faced by \acp{UAV} during navigation can vary significantly from one environment to another, making the applicability of \ac{DQL} solutions limited in scope. 

Compared with these prior works, we do not focus on designing a more advanced \ac{RL} algorithm to improve the performance of the system. Instead, we investigate how to share prior model knowledge to improve the \ac{RL} training convergence speed when the \ac{UAV} faces a new environment.
\ac{TL} has emerged as an effective approach among researchers to address this challenge. \cite{chen2020knowledge} applied \ac{TL} to enhance tracking performance by learning from the tracking errors of \acp{UAV} with different dynamics without requiring baseline controller modifications. Furthermore, \cite{fontanesi2022transfer} utilized the teacher policy trained in a sub-6 GHz domain, which accelerates path learning in a new \ac{mmWave} domain. The authors also considered outage constraints and used a robust \ac{DDQN} as the base model. 
In this study, a pre-trained model involves initially training a \ac{DDQN}, but targeting different domains that facilitate the transfer of knowledge across various environments. In \cite{zhang2020trajectory}, author utilizes \ac{TL} to adapt \ac{UAV} trajectory design to emergency scenarios where user distribution and terrain change. Mainly simulate the scenario of certain \acp{BS} losing their functions in emergency situations. However, this method falls short of addressing the challenge of knowledge transfer across diverse urban settings. Thus, it's crucial to explore the capability for continuous learning across various environments, beyond just adapting to situational and task changes within a single map.

\vspace{-0.1in}
\subsection{Contributions}
This article is dedicated to evolving \ac{UAV} path planning training models into adaptable frameworks that facilitate faster and repeated retraining in different environments with \ac{CTL}. 
We propose a shift from the vanilla approach of one-time model training to a paradigm of continuous adaptation, where \acp{UAV} are retrained with new data and conditions as they transition between tasks and environments. While an ad hoc trained \ac{RL} model can only be applied to a specific environment, this approach is not limited to a specific environment and provides a flexible solution for different scenarios.
We propose a \ac{CTL} framework to enable \ac{RL} for \acp{UAV} trajectory with connectivity constraint to rapidly adapt to different environments.  
This approach involves first pre-training a model to achieve a specific convergence target, serving as a foundational base to reduce the cost of training the model in the new environments. To this aim, we train a policy for \ac{UAV} trajectory that tackles ground-to-air link outages and we evaluate both \ac{DQN} and \ac{DDQN} models. Among various models with differing hyper-parameters, the higher-performing trained \ac{DDQN} is chosen as the foundational learning model. A detailed analysis is provided in Section III. Then, our method involves transferring the learned model weights and adjusting the learning parameters, including the learning rate and exploration rates, to the specific characteristics of each new environment.
Specifically, we consider the problem space formed by three scenarios, namely: Environment 1 (dense urban landscape with tall buildings), Environment 2 (emergency scenario), Environment 3 (suburban residential area). We evaluate the effectiveness of our \ac{CTL} approach through various tasks, targeting different destinations, and in emergency conditions involving \acp{BS} failures. Simulation results show significant improvements in both convergence times and stability in new environments.

\vspace{-0.1in}
\section{System Model and Problem Formulation}
\subsection{System Model}
\begin{figure*}[t]
     \centering
     \begin{subfigure}[b]{0.32\textwidth}
         \centering
         \includegraphics[width=\textwidth]{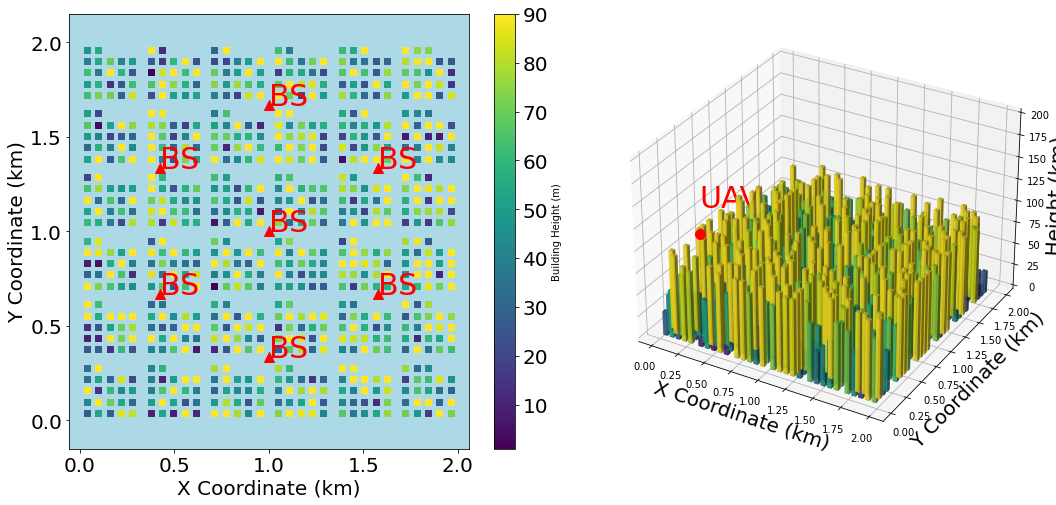}
         \caption{Environment 1}
         \label{fig:Environment_1}
     \end{subfigure}
     \hfill
     \begin{subfigure}[b]{0.32\textwidth}
         \centering
         \includegraphics[width=\textwidth]{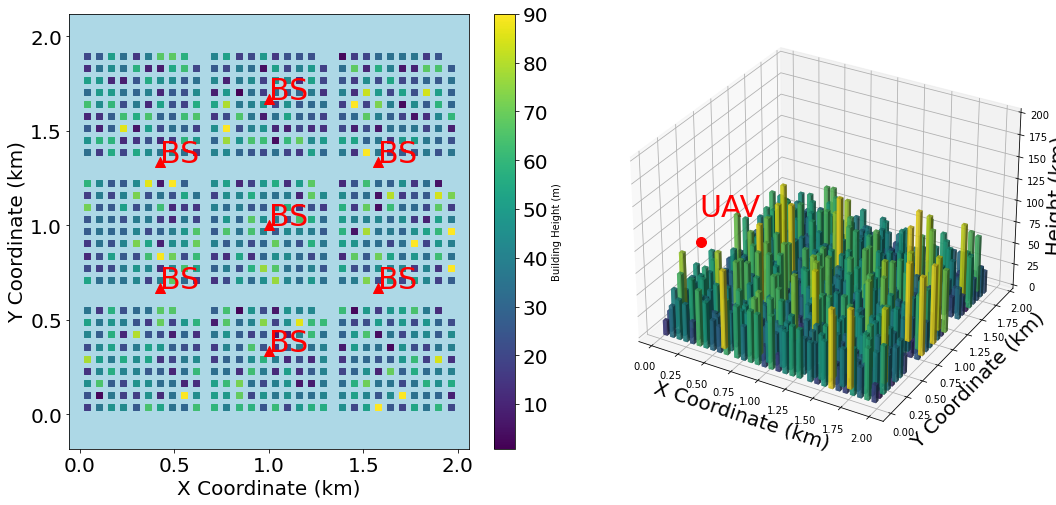}
         \caption{Environment 2}
         \label{fig:three_sinx}
     \end{subfigure}
     \hfill
     \begin{subfigure}[b]{0.32\textwidth}
         \centering
         \includegraphics[width=\textwidth]{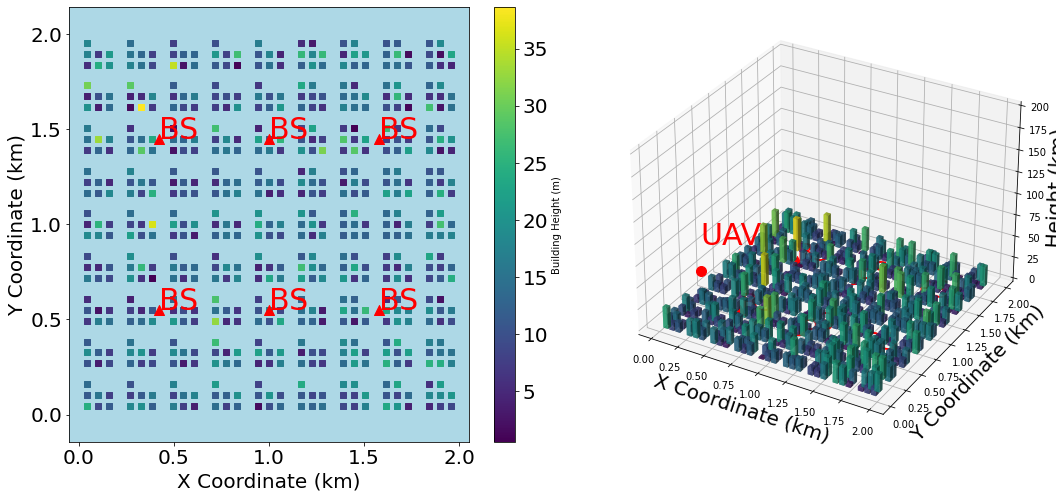}
         \caption{Environment 3}
         \label{fig:five_over_x}
     \end{subfigure}
         \caption{Comparative Visualization of UAV Operational Environments: This figure illustrates the UAV's navigation challenges and strategies across three environments. Graph a depicts a dense urban landscape with tall buildings; Graph b shows a sparse urban area with two scenarios: a standard mission and an emergency scenario, emphasizing the UAV's adaptability to sudden environmental changes; Graph c shows a third, more differentiated scenario, modelling a suburban residential area and varying the distribution of base stations}\small
         \vspace{-0.1in}
        \label{fig:three_graphs}
\end{figure*}
We consider a singular \ac{UAV}-aided cellular network.
The core objectives of \ac{UAV} are to navigate swiftly and efficiently toward a designated target location while ensuring seamless communication with a terrestrial cellular network. This operational versatility extends to diverse geographical regions, denoted as space of Environments $\mathcal{S}$ ($X_1, X_2, X_3$), as shown in Fig. \ref{fig:three_graphs}, where each region presents unique building distribution and communication needs. Map distinctions between different environments include factors such as urban building type, building density, street width, base station height, and other relevant parameters. 

\acp{BS} are deployed at specific locations within the urban area. Let \( \mathcal{M} \) be the set of \acp{BS}. Each \ac{BS} \( m \in \mathcal{M} \) is characterized by its geographical location \((x_s, y_s)\) and height \(h_s\) with 3 sectors $j $. The \ac{UAV} embarks on its mission from a designated initial location, represented as $q_I \in \mathbb{R}^{3\times1}$, which can vary for each deployment. The primary goal is to navigate the \ac{UAV} to a predetermined target point, marked as $q_F \in \mathbb{R}^{3\times1}$, while ensuring uninterrupted communication links with the terrestrial network during the mission, with the probability of communication outage maintained below requirement.
The \ac{UAV} moves at constant speed $V=V_{max}$ along a 3D trajectory of duration $T$ that can be divided into $K$ discrete segments with interval $\delta_k = T/K$, $k = \{1,...,K\}$.  $\delta_k$ is chosen arbitrarily small so that within each step the large scale signal power received by the UAV remains approximately unchanged. Each segment is thus described by its discrete coordinates $\mathbf{q}(n)= (x_n, y_n, h_n)$.

The channel model considers various factors, including large-scale path loss, small-scale fading, and environmental elements like terrain and buildings. Large-scale path loss represents the signal attenuation over distance and frequency. The path loss is typically classified into: LoS (Line-of-Sight) Path Loss Model: $l_L(d) = X_L \times d^{-\alpha_L}$. And NLoS (Non-Line-of-Sight): $l_{NL}(d) = X_{NL} \times d^{-\alpha_{NL}}$ \cite{fontanesi2020outage}. We utilize the three-sector antenna model (downtilted to serve ground \acp{UE} \cite{fontanesi2022transfer}), as defined by the 3rd Generation Partnership Project (3GPP) specifications \cite{mondal20153d}. The antenna gain is a critical component of signal reception from ground \ac{BS} to a \ac{UAV}, which can be represented as $G_{m,j}(\theta, \phi) = A_{3GPP E}(\theta, \phi) + AF(\theta, \phi, n)$. It combines the 3GPP antenna element pattern ($A_{3GPP E}(\theta, \phi)$) and array factor ($AF(\theta, \phi, n)$ to represent beamforming effects \cite{rebato2018study}.

Fading phenomena describe signal variations over time due to environmental changes, particularly \ac{UAV} movement. Fading coefficients $f(t)$ represent instantaneous fading levels, while small-scale fading powers $f^2_{0,i}$ for both LoS and NLoS environments follow Nakagami-M fading models \cite{fontanesi2020outage}.
To compute the Signal-to-Interference-plus-Noise Ratio (SINR) at a UAV from a specific ground \ac{BS} ($m$) sector ($j$), the Signal Power $P_s = P_{t,m} \times G_{m,j}(\theta, \phi) \times L_{path}(d) \times L_{fading}$, Interference Power $P_i = \sum_{i \neq m} P_{t,i} \times G_{i}(\theta_i, \phi_i) \times L_{path}(d_i) \times L_{fading_i}$, and Noise Power $P_n = N_0 \times B$ are considered. $P_s$ accounts for the desired signal power. $P_i$ aggregates interference from all other \acp{BS}, and $P_n$ represents noise power. The SINR is computed as $SINR_{max,m,j} = \frac{P_{s,j}}{P_i + P_n}, (j \in J = {1, 2, 3})$.\\
During \ac{UAV}'s path, we consider an outage occurs when the SINR is less than or equal to $\varphi_{\text{th}}$, leading to an outage events: $\beta(\mathbf{q}(n)) = 1$ if $\left(\text{SINR}(q(n)) \leq \varphi_{\text{th}}\right)$. The total outage events is denoted by $\Gamma = \sum_{n=0}^{T} \beta(\mathbf{q}(n)) \,$.



\subsection{Problem Formulation}
The objective of this research is to improve the adaptability of \ac{RL} models, specifically \ac{DDQN}, to novel environments using \ac{CTL}. 


Our general goal is to ensure that the training time for a policy $\pi^s$ in Environment $s \in \mathcal{S}$ is significantly less than the time required to train a new model from scratch for each new environment. 
We can thus formulate
\begin{subequations}\label{eq:Opt_problem1}
    \begin{align}
  \mathcal{P}:  \min_{} &\quad  T_{\text{train}}(\pi^s)\\ \label{eq:cond1_Opt_problem1}
    \textrm{s.t.} \quad & \pi_{\text{new}}^{(0)} = \pi_{\text{pre-trained}},
    \end{align}
\end{subequations}
where denoted by $T_{\text{train}}(\pi^s)$ is the training time of policy $\pi^s$ in the new enviroment $s$ and \eqref{eq:cond1_Opt_problem1} specifies that a policy $\pi^1$ pre-trained in Environment 1 $\in \mathcal{S}$ is used as benchmark for the \ac{UAV} trajectory in the new environment.
Policy $\pi^1$ is trained using a \ac{DDQN} model and then utilized for transfer learning across two additional maps or scenarios. This approach necessitates that after the transfer, the model maintains its original trajectory objectives, i.e., minimizing the steps $n$ needed to reach its destination and to keep the frequency of outage events $\Gamma$ below a certain threshold $\Hat{\Gamma}$. Policy $\pi^1$ is thus trained to solve problem:
\begin{align}
\vspace{-0.1in}
\min_{n, q(n)}  &\quad    n + K \times \Gamma   \\ 
\text{s.t.} \quad & \mathbf{q}(0) = q_I, \quad \mathbf{q}(T) = q_F \\ \label{eq:cond2_Opt_problem2}
& \Gamma < \Hat{\Gamma}   \\
& h(\text{u}) > h_{\text{B}} \\
& n \leq N.
\end{align} 
%
The variable \(q(n)\) represents the \ac{UAV}'s position at step \(n\), \(q_I\) and \(q_F\) signify the initial and final positions of the \ac{UAV}, while q(T) presents the target position. 
Constraint \eqref{eq:cond2_Opt_problem2} ensures that the maximum outage events throughout the flight remain below the specified threshold. Here \(h(\text{u})\) denotes the altitude of the UAV, \(h_{\text{B}}\) represents the highest altitude of the building, and \(N\) signifies the maximum permissible movement step of the \ac{UAV}, considering battery constraints.

\section{RL and Continuous Learning for UAV Trajectory Design}
\subsection{Preliminaries}

\ac{DQN} and its enhancement, Double \ac{DQN}, represent significant advancements in integrating reinforcement learning with deep neural networks for autonomous decision-making. \ac{DQN} facilitates learning from complex sensory inputs, overcoming challenges like instability and convergence issues through techniques such as experience replay and fixed Q-targets. \ac{DDQN} further refines this by correcting \ac{DQN}'s overestimation of Q-values, ensuring more stable and accurate outcomes. This progression makes \ac{DQN} and \ac{DDQN} highly effective for \ac{UAV} trajectory optimization in diverse urban environments, improving navigation precision in complex scenarios.

\ac{TL} further enhances the learning process by utilizing knowledge acquired in one task to expedite learning in related but distinct tasks. It operates on the principle of reusing a pre-trained model as a starting point for new tasks, significantly cutting down on the time and data needed for training in new environments or missions. Continuous Learning complements these methodologies by enabling \acp{UAV} to continually assimilate new information without discarding previously learned knowledge. This is crucial for operating in dynamic environments where conditions and obstacles may change unpredictably. Continuous learning ensures that \acp{UAV} can iteratively update their navigation strategies, maintaining peak performance and adaptability over time.

\vspace{-0.1in}
\subsection{ DQN and DDQN for UAV Trajectory Design}

In our considered scenario, the \ac{UAV} navigation task is modeled in a 3D space, where the \ac{UAV} must reach a target position, starting at a random position within the set range. Given our \ac{UAV} trajectory optimization, we can formulate the problem as an \ac{MDP}, which is defined by a tuple \( ( \mathcal{S}, \mathcal{A}, \mathcal{P}, \mathcal{R}, \gamma) \). Each state \( s \) represents the UAV's current position, communication conditions, and the environment. Formally:
\[ s = \{ q(n), \text{SINR}(q(n))\}, \]
where \( q(n) \) and \( \text{SINR}(q(n)) \) are the \ac{UAV}'s position and signal quality, respectively. The action space comprises potential directions in which the \ac{UAV} can move at any given time:\[ \mathcal{A} = \{ \text{move\_f}, \text{move\_b},\text{move\_l}, \text{move\_r} \}. \] Given a state \( s \) and an action \( a \), \( \mathcal{P}(s'|s, a) 
\) represents the probability of transitioning to state \( s' \). Since the \ac{UAV}'s movement is deterministic, given its current state and action, it will deterministically arrive at the next state. The reward function captures the objectives of minimizing the completion time and maintaining satisfactory communication. It is presented as
\vspace{-0.08in}
\begin{equation}\label{eq:Opt_problem1}
\mathcal{R}_{1}(s, a, s') = -k1 \times d(s, s_{\text{target}}) - k2 \times F(q_n) - R_{\text{n}} + R_{\text{arrive}}
\vspace{-0.in}
\end{equation}
where k1 and k2 are weighting factors to prioritize moving closer to the destination and penalize outage events during the transition. \( d(s, s_{\text{target}}) \) represents the distance between the \ac{UAV}'s current position and the target position. \( F(q_n) \) denotes the signal outage penalty function, which increases the loss for navigating through areas with poor signal quality. $R_{\text{n}}$ is the step penalty, and $R_{\text{arrive}}$ is the reward for reaching the target. The discount factor \( \gamma \) determines the present value of future rewards. Given the \ac{MDP} framework for the UAV trajectory optimization, our goal is to find a policy \( \pi^* \) that maximizing an expected cumulative reward:
\vspace{-0.10in}
\begin{equation}
\pi^* = \arg\max_{\pi} \mathbb{E}\left[\sum_{t=0}^{T} \gamma^t \mathcal{R}(s_t, a_t, s_{t+1})\right],
\end{equation}
where $ \mathbb{E} $ denotes the expectation. The \ac{DQL} agent uses two neural networks: a primary network for action selection and a target network for stable learning. Both \ac{DQN} and \ac{DDQN} utilize a similar network structure for action selection and evaluation, but \ac{DDQN}'s critical distinction lies in its action evaluation mechanism, which separates action selection from its value estimation. This separation ensures a more conservative and accurate assessment of the potential of each action, thereby enhancing the performance of trajectory design. In addition to this, we have designed \ac{MDP} that is more compatible with our scenarios to improve performance. This includes adding SINR information to the state to ensures that the models for transfer have sufficient depth of learning and understanding of the environment. The rewards will also be fine-tuned according to the situation when migrating to a new environment.

\vspace{-0.1in}
\subsection{Continuous Learning for UAV Trajectory Design}

\subsubsection{Environments}
We define the first environment that was used to train the \ac{DDQN} base model as a dense urban area, characterized by a high concentration of tall buildings that demand complex navigational strategies to maintain connectivity and avoid obstacles. Transitioning to the second environment, the \ac{UAV} encounters a sparse urban landscape with shorter buildings, a stark contrast to the first. This environment is explored under two distinct scenarios which are a standard mission mirroring the parameters of the dense urban environment, and an emergency scenario necessitating a sudden change in mission objectives due to a \ac{BS} failure. The adaptability of the \ac{UAV} is further tested in the third environment, a residential area with low housing blocks and a different \ac{BS} distribution, presenting new navigational challenges and testing the \ac{UAV}'s ability to generalize its learned strategies to markedly different landscapes.
\subsubsection{ Training in different Domain}
\begin{algorithm}
\caption{\ac{CTL} with DDQN across Environments} \label{algo:2}
\begin{algorithmic}[1]
\State Initialize DDQN in Environment 1: learning rate $\alpha_1$, exploration $\epsilon_1$, and exploration decay $\epsilon_{\text{decay},1}$.
\State Train the model with a replay buffer and a periodically updated target network to stabilize updates.
\State Save the model weights $\xi_{\text{Env1}}$ and policy $\pi_{\text{Env1}}$
\State \textbf{Transfer to Env2:}
\State Initialize Env2 with DDQN model weights $\xi_{\text{Env2}} \leftarrow \xi_{\text{Env1}}$ and policy $\pi_{\text{Env2}} \leftarrow \pi_{\text{Env1}}$
\State Adjust learning parameters for Env2: set new learning rate $\alpha_2$, $\epsilon_2$, and $\epsilon_{\text{decay},2}$
\State Apply the new reward function \({R}_{2}\)(optional).
\State Retrain transferred model in Env2 using the fixed parameters, maintaining the strategy of experience replay and target network updates.
\State Save the updated model weights $\xi_{\text{Env2}}$ and policy $\pi_{\text{Env2}}$
\State \textbf{Transfer to Env3} with process above
\State Continue this process for any subsequent environments, transferring weights and policy while adjusting learning parameters as needed
\end{algorithmic}  
\end{algorithm}
The \ac{TL} process begins with training a \ac{DDQN} model in a dense urban setting, streamlining the training approach to enhance efficiency without requiring full model convergence. A targeted termination criterion focuses on achieving a high success rate at the destination rather than optimizing the total reward, shortening training by 600-800 episodes. This method is chosen because, after reaching success rate convergence, the agent's main improvements involve strategies to avoid outage events, which may vary in effectiveness across different environments. This efficient training strategy reduces overall duration and improves the model's suitability for \ac{CTL}. Subsequent findings affirm the benefits of this balanced approach, demonstrating its practical success.

This pre-trained model is then transferred to accommodate sparse urban and residential areas. This process entails transferring the learned model weights and adjusting learning parameters, including the learning rate and exploration rates, to suit each new environment's specific characteristics (algorithm \ref{algo:2}). Through this methodical adaptation, the \ac{UAV} demonstrates not only enhanced learning efficiency but also improved performance and adaptability across diverse operational environments. By leveraging the knowledge gained in each successive environment, \ac{CTL} enables the \ac{UAV} to rapidly adjust to new challenges. The objective in the new environment remains aligned with our dual goals. The model, now starting with policy \( \pi_{\text{new}}^{(0)} \), is further trained to adapt to the new conditions. The fine-tuning process involves iterative updates of the policy parameters:
\begin{align}\label{eq:learningRate}
\vspace{-0.3in}
\pi_{\text{new}}^{(0)} &= \pi_{\text{pre-trained}}, \\
\pi_{\text{new}}^{(k+1)} &= \pi_{\text{new}}^{(k)} - \alpha \cdot \nabla_{\pi_{\text{new}}} \mathcal{L}(\pi_{\text{new}}^{(k)})
\end{align}
where \( \alpha \) is the learning rate, and \( \mathcal{L} \) represents the loss function tailored to the new environment.

Then we evaluate the performance of the transferred model in the new environment using metrics such as convergence time \( T_{\text{convergence}} \) and improvements in stability or accuracy. These metrics will demonstrate the effectiveness of \ac{TL} in reducing training time and enhancing model performance in diverse urban scenarios.
%
%
%
\begin{table}[t]
\centering
\caption{Summary of Model Parameters}
\label{tab:parameters}
\begin{tabular}{lll}
\toprule
Parameter Name & Symbol & Value \\
\midrule
Learning Rate(DDQN)& \( \alpha \) & 0.001 \\
Initial Exploration Rate(DDQN)& \( \epsilon \) & 1.0 \\
Exploration Decay Rate(DDQN) & \( \epsilon_{\text{decay}} \) & 0.998 \\
Learning Rate (Transfer) & \( \alpha_{\text{transfer}} \) & 0.0002 \\
Initial Exploration Rate (Transfer) & \( \epsilon_{\text{transfer}} \) & 0.5 \\
Exploration Decay Rate (Transfer) & \( \epsilon_{\text{decay, transfer}} \) & 0.995 \\
Arrive Target for Env1 & \( q_F1 \) & 1000,900 \\
Arrive Target for Env2  & \( q_F2 \) & 1250,1300 \\
Discount Factor & \( \gamma \) & 0.95 \\
Weighting Factor for Distance length& \( k_1 \) & 0.8 \\
Weighting Factor for Outage Penalty & \( k_2 \) & 1 \\
Step Penalty & \( R_n \) & 1\\
Reward for Reaching Target & \( R_{\text{arrive}} \) & 2000 \\
Max steps per eposode & \( \text{steps}\) &200\\

SINR Threshold for outage& \( \varphi_{\text{th}} \) & 0 dB \\
UAV Height & \( h(\text{u}) \) & 90 m \\
\bottomrule
\end{tabular}
\vspace{-0.1in}
\end{table}
\begin{figure}[t]
    \centering
    \includegraphics[width=1\columnwidth]{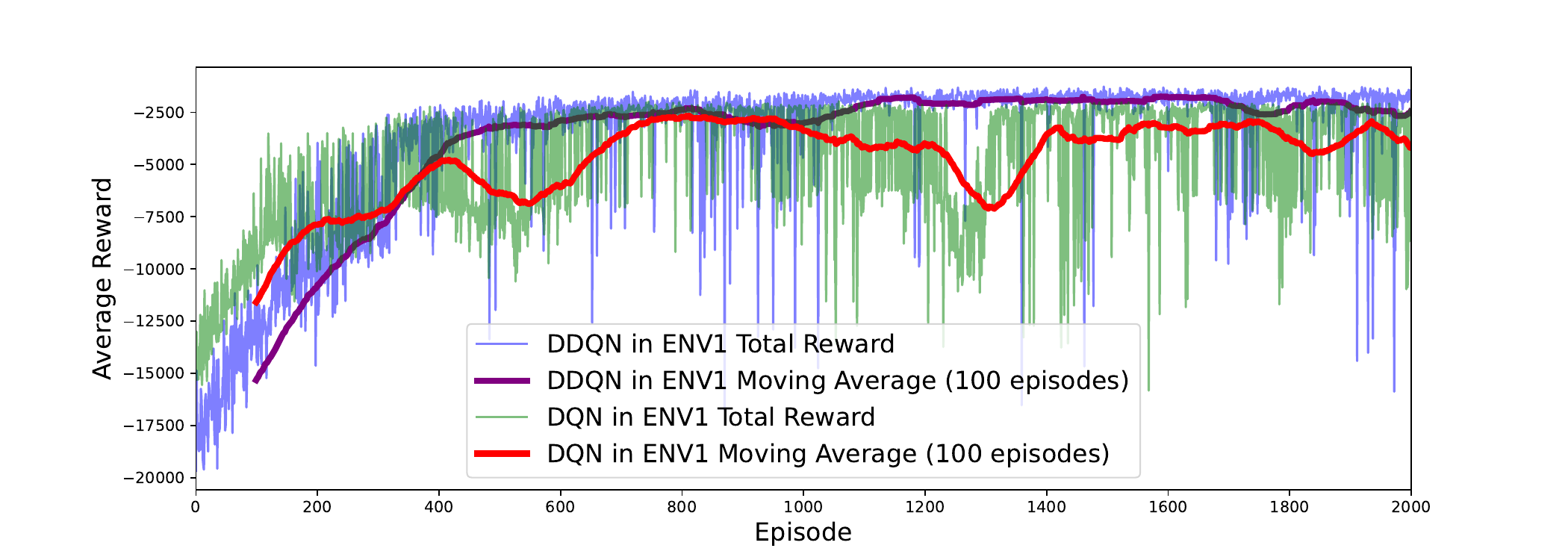}
    \caption{Compare of DQN with DDQN for Trajectory Design in Env1}\small
    \label{fig:1.png}
    \vspace{-0.1in}
\end{figure}
\vspace{-0.1in}

\section{Results}
The height of the \ac{BS} on the horizontal plane ranges from 5 to 25 meters, reflecting typical real-world configurations. The operational area for all maps is set to a size of $2 \, \text{km} \times 2 \, \text{km}$ with a height limit of 100 meters. The maximum number of steps per episode, representing the battery limitation, is set to 200 steps, with each step corresponding to a displacement of 10 meters. A \ac{UAV} arriving within a 30 meters radius of the target is considered to have successfully reached its destination. 
The networks consist of 3 hidden layers with 64 units each, using ReLU activation functions. The output layer has linear activation corresponding to action values. The learning rate in \eqref{eq:learningRate} was reduced to 0.0002 for fine-tuning, allowing for more subtle weight adjustments in the later layers, thereby refining the model's policy without drastic deviations from its pre-learned behaviors. Weighting factors for rewards \eqref{eq:Opt_problem1} and more training parameters can be found in Table \ref{tab:parameters}.

The agent's performance improved over time, as indicated by an increase in total reward and a decrease in the number of steps required to reach the target. The learning progress was captured through two plots: total rewards, success rate over episodes, and average level of communication.  
To demonstrate this, we compared how well \ac{DQN} and \ac{DDQN} adapted to our scenarios, choosing the one that performed better as our base model for continuous learning.

In Fig. \ref{fig:1.png} we illustrate the comparative performance between the \ac{DQN} and \ac{DDQN} models within the initial environment. The \ac{DQN} model exhibits greater volatility throughout the training phase, with a 10-15\% lower peak in average rewards for the optimum policy compared to the \ac{DDQN}. Additionally, the \ac{DQN} demonstrates less stable convergence by the 2000 episode, accompanied by increased variability. Thus, the \ac{DDQN} model emerges as a more suitable foundational model for \ac{TL}. 
\vspace{-0.1in}
\subsection{CTL in environment 2 with multi scenarios}

\begin{figure}[h]
  \centering
  \begin{subfigure}[b]{0.52\textwidth}
    \includegraphics[width=\textwidth]{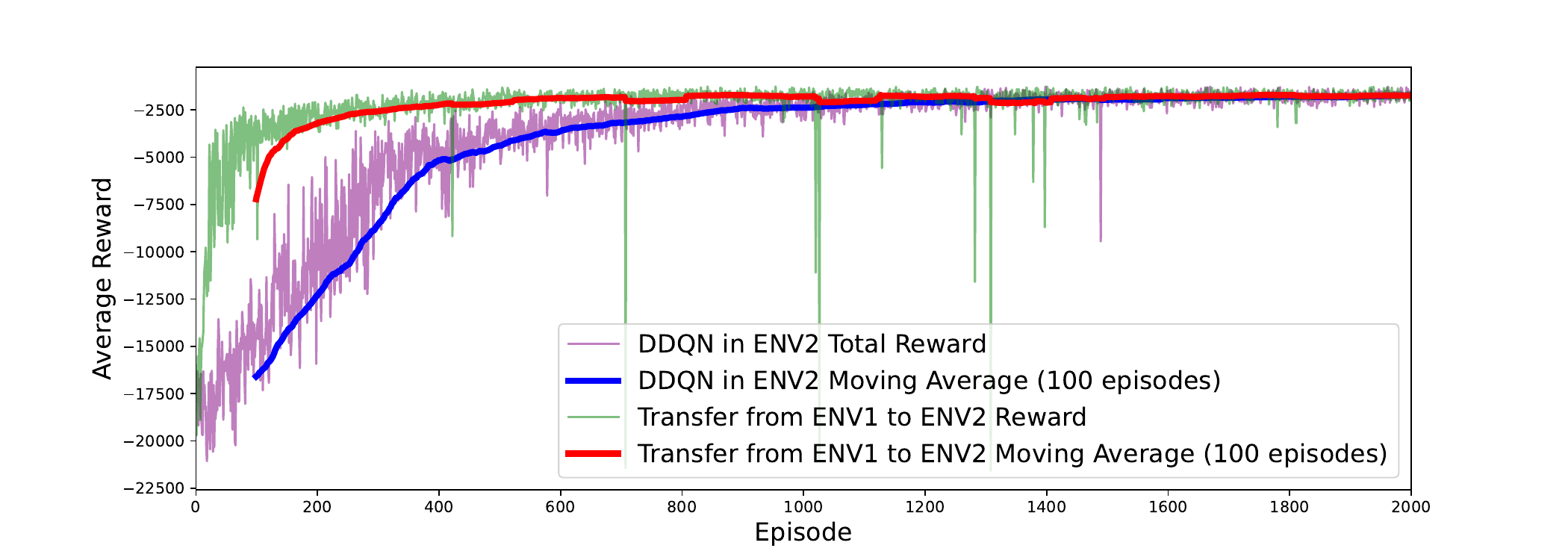}
    \caption{Average rewards against episodes}
    \label{fig:TransferLearning_comparisonA}
  \end{subfigure}
  \hspace{5pt} %
  \begin{subfigure}[b]{0.52\textwidth}
    \includegraphics[width=\textwidth]{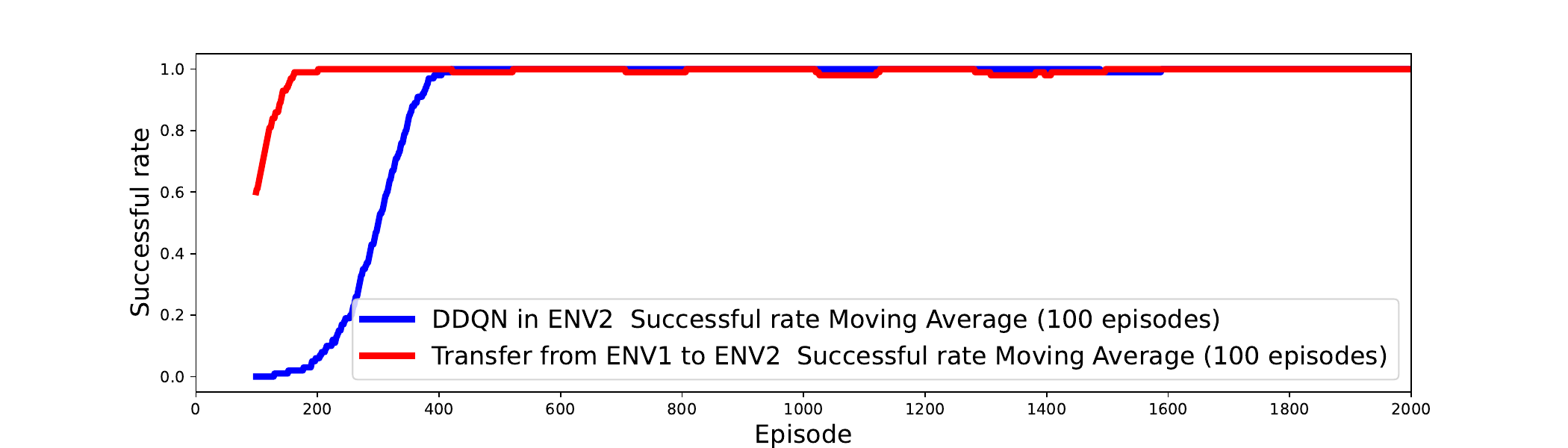}
    \caption{Success rate against episodes}
    \label{fig:TransferLearning_comparisonB}
  \end{subfigure}
    \caption{\ac{TL} from ENV1 to ENV2 against retraining DDQN in ENV2}\small
    \label{fig:TransferLearning_comparison}
\end{figure}

\begin{figure}[t]
  \centering
  \begin{subfigure}[b]{0.5\textwidth}
    \includegraphics[width=\textwidth]{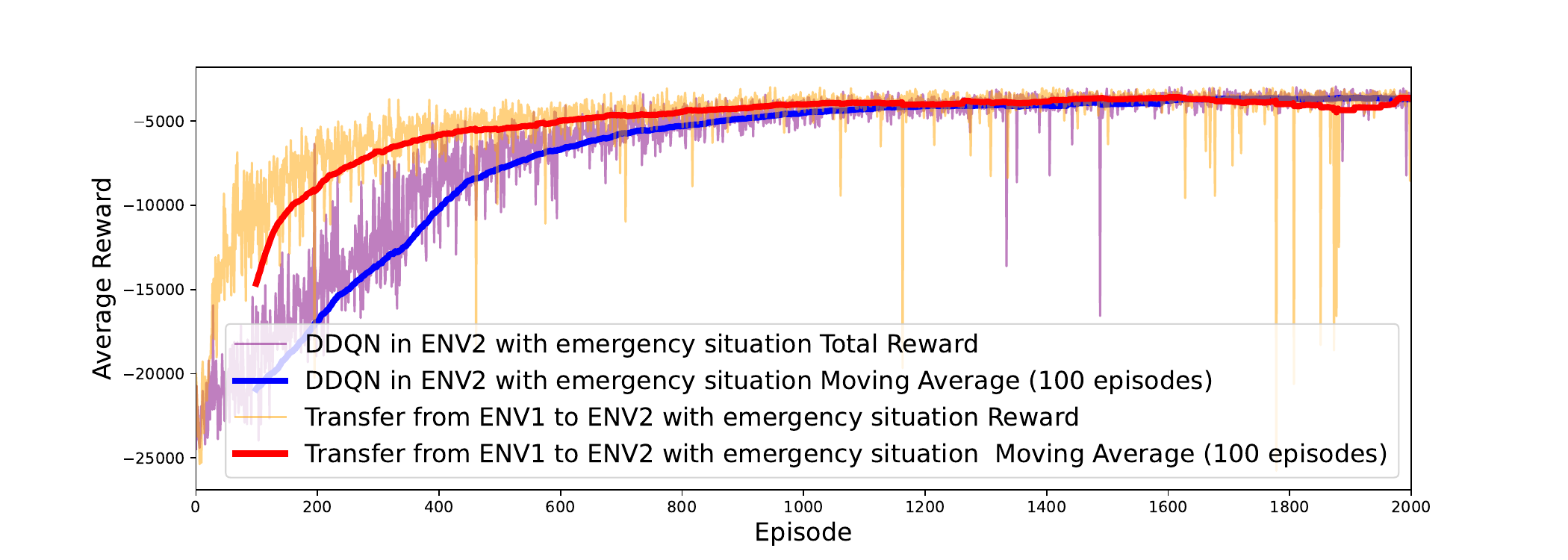}
    \caption{Average rewards against episodes }
    \label{fig:TransferLearning2.2R}
  \end{subfigure}
  \hspace{5pt} %
  \begin{subfigure}[b]{0.5\textwidth}
    \includegraphics[width=\textwidth]{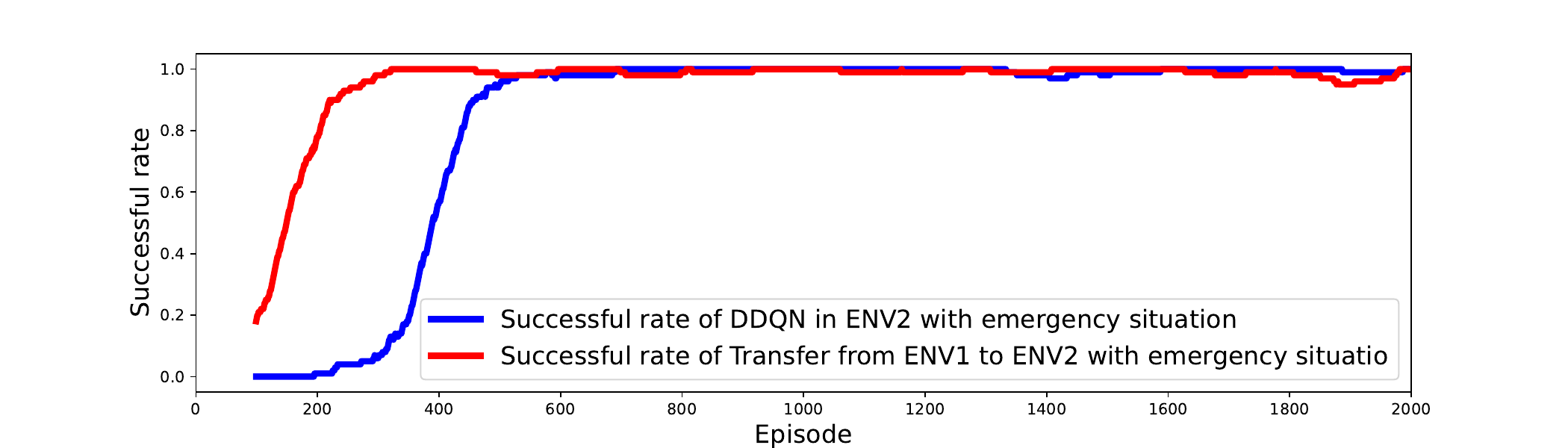}
    \caption{Success rate against episodes}
    \label{fig:TransferLearning2.2S}
  \end{subfigure}
    \caption{\ac{TL} from ENV1 to ENV2 with emergency scenario against retraining \ac{DDQN} in ENV2}
    \label{fig:TransferLearning_comparison2}\small
    \vspace{-0.2in}
\end{figure}

This section shows the outcomes of \ac{TL} in the second environment, encompassing two scenarios. Scenario one, illustrated in Fig. \ref{fig:TransferLearning_comparison}, involves utilizing the \ac{DDQN} model from the first environment as the foundational model and has the same mission destinations as \ac{DDQN}. The result in Fig.\ref{fig:TransferLearning_comparison2} shows a more challenging scenario, which has different target positions and one of the \ac{BS} is not working.

Fig.\ref{fig:TransferLearning_comparisonA} demonstrates the role of \ac{TL} by showing the average rewards of the training, showcasing an earlier and more stable convergence compared to the \ac{DDQN} model trained from scratch. The duration to achieve stable optimal rewards is shortened by 600–700 episodes, underscoring the efficiency of \ac{TL} in accelerating performance and stability. The success rate of reaching the destination during training also notably underscores the efficiency of \ac{TL}. As shown in Fig.\ref{fig:TransferLearning_comparisonB}, the success rate stabilizes at least 99\% over 250 episodes sooner than the scratch-trained model, which means that the model prioritizes finding the policy that reaches the destination and then adapts the relevant policy for the communication much faster, and this process for \ac{DDQN} is much slower.

In navigating the challenges of training in a environment where a base station BS was no longer functional and the target position was further away, the \ac{UAV} demonstrated commendable adaptability. Fig.\ref{fig:TransferLearning2.2R} illustrates the average rewards of \ac{TL}, and Fig.\ref{fig:TransferLearning2.2S} shows the success rate in Environment 2 where, despite facing significant challenges, \ac{TL} demonstrates its effectiveness by converging 300 episodes earlier in terms of average reward and 200 episodes earlier in reaching the target than the model trained from scratch. This scenario limited the \ac{CTL} approach from fully exploiting all initial strategies as compared to a model beginning from scratch. It shows slight fluctuations during the training phase, these did not hinder the overall process convergence, which showcases the robustness of the \ac{UAV}'s learning capabilities. The observed variations, partly due to the task's target location moving further away, marginally slowed the learning speed but did not detract from the \ac{UAV}'s ability to readjust and progress. Therefore, the performance of \ac{CTL} is expected and acceptable when faced with more challenging scenarios and altered tasks. The \ac{CTL} efficiently navigated these complexities, underlining the effectiveness of its adaptive learning framework in dynamic environments.

\vspace{-0.1in}

\subsection{\ac{CTL} in environment 3 more differences}

\begin{figure}[h]
  \centering
  \begin{subfigure}[b]{0.5\textwidth}
    \includegraphics[width=\textwidth]{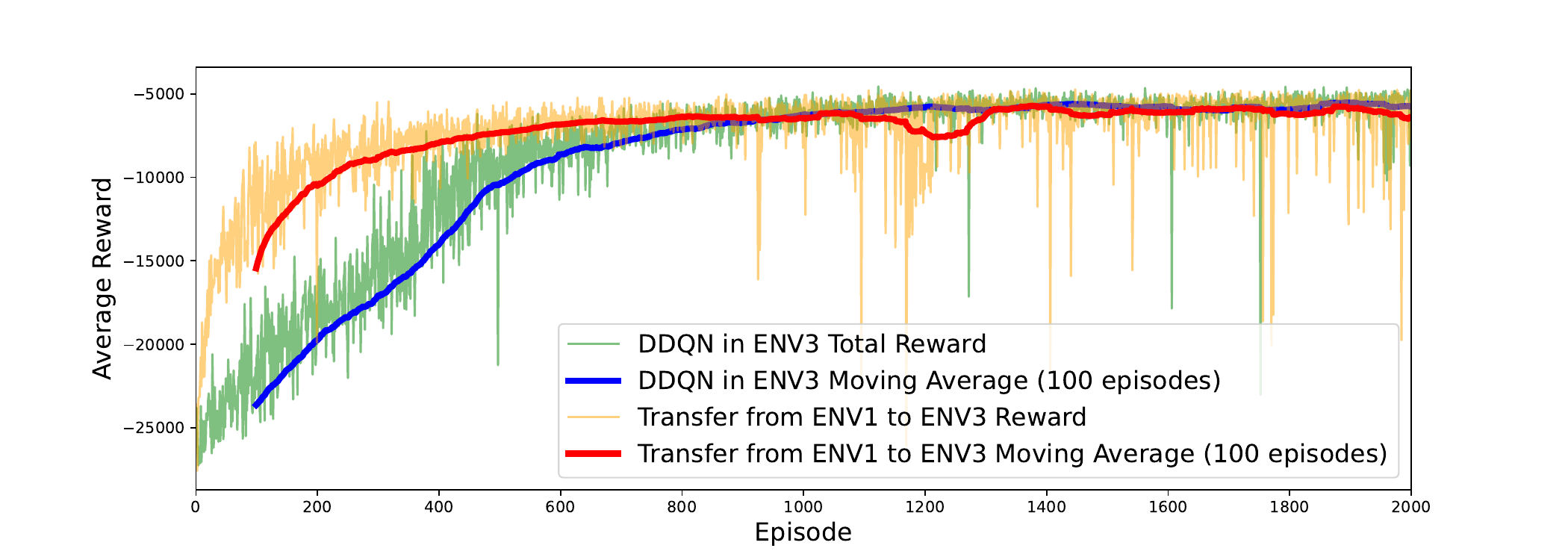}
    \caption{Average rewards against episodes  }
    \label{fig:1to3.png}
  \end{subfigure}
  \hspace{5pt} %
  \begin{subfigure}[b]{0.5\textwidth}
    \includegraphics[width=\textwidth]{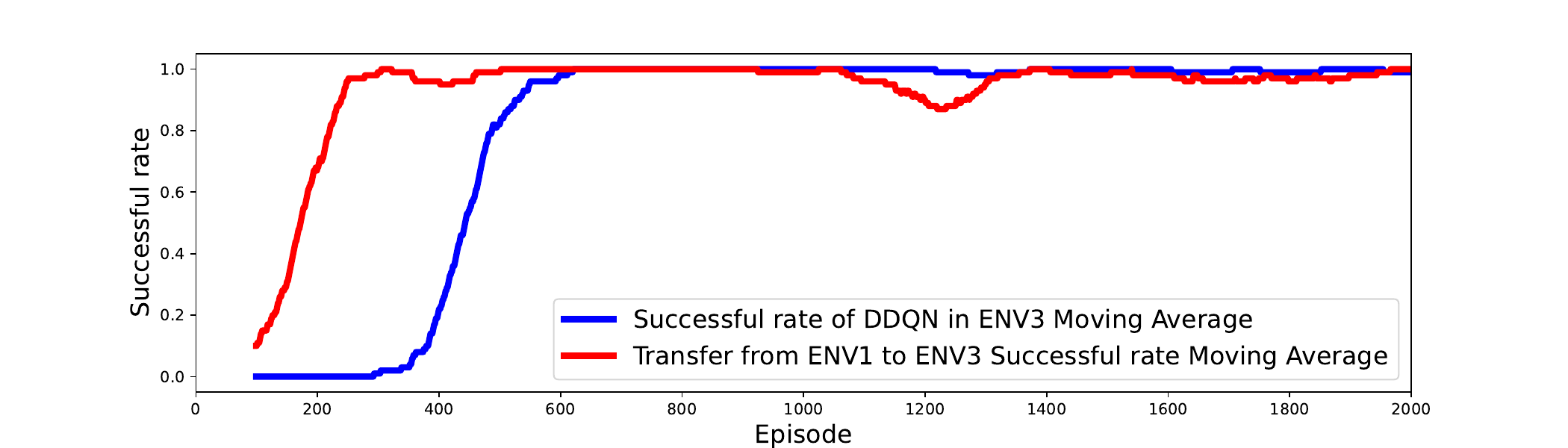}
    \caption{Success rate against episodes }
    \label{fig:1to3srate}
  \end{subfigure}
    \caption{\ac{TL} from ENV1 to ENV3 against retraining DDQN in ENV3}\small
    \label{fig:TransferLearning3}
    \vspace{-0.2in}
\end{figure}

In the third environment (Fig.\ref{fig:TransferLearning3}), the role is to test the potential for \ac{CTL}. We designed the third environment to be very different from the environment used for the base model. This includes the distribution of buildings, height, size, number of \ac{BS} locations, and changes in the target destination. The \ac{CTL} still demonstrates strong capabilities; the convergence was 200 episodes faster and reached a stable success rate 150 episodes earlier. However, more fluctuations were encountered later on, which is predictable for reasons similar to those discussed in Environment 2, as not all environments were explored completely. This serves as an effective illustration of the capabilities of continuous learning to expedite model training outcomes, even when faced with environments that present significant disparities.

\vspace{-0.1in}
\section{Conclusion}

In conclusion, the performance of \ac{CTL} across two distinct environments demonstrated its effectiveness in accelerating learning and achieving higher success rates compared to training \ac{DDQN} models from scratch. For transfer to similar environments, \ac{TL} showcased a notable advantage in terms of stability and early convergence, with the second case highlighting its capacity to adapt to more complex tasks and scenarios, albeit with reduced efficiency and some late-stage fluctuations. The third environment, significantly divergent from the initial training context, further tested the limits of \ac{CTL}, where, despite faster convergence and commendable performance, the model faced increased fluctuations due to incomplete exploration of the new environment. These findings underscore the potential of \ac{TL} to enhance model adaptability and efficiency, particularly in dynamically changing or progressively complex scenarios, while also pointing to the need for strategies to mitigate late-stage performance variability. In our future works we plan to investigate this issue and work on strategies that deal with this performance variability, and more in-depth study of energy conservation strategies.

\begin{acronym} 
\acro{5G}{Fifth Generation}
\acro{6G}{Sixth Generation}
\acro{AI}{Artificial Intelligent }
\acro{ACO}{Ant Colony Optimization}
\acro{ANN}{Artificial Neural Network}
\acro{BB}{Base Band}
\acro{BBU}{Base Band Unit}
\acro{BER}{Bit Error Rate}
\acro{BS}{Base Station}
\acro{BW}{bandwidth}
\acro{C-RAN}{Cloud Radio Access Networks}
\acro{CU}{Central Unit}
\acro{RU}{Radio Unit}
\acro{DU}{Distributed Unit}
\acro{CAPEX}{Capital Expenditure}
\acro{CoMP}{Coordinated Multipoint}
\acro{CR}{Cognitive Radio}
\acro{CRLB}{Cramer-Rao Lower Bound}
\acro{C-RAN}{Cloud Radio Access Network}
\acro{CTL}{Continuous Transfer Learning}
\acro{D2D}{Device-to-Device}
\acro{DAC}{Digital-to-Analog Converter}
\acro{DAS}{Distributed Antenna Systems}
\acro{DBA}{Dynamic Bandwidth Allocation}

\acro{PID}{Proportional–integral–derivative}

\acro{DC}{Duty Cycle}
\acro{DFRC}{Dual Function Radar Communication}
\acro{DL}{Deep Learning}
\acro{DSA}{Dynamic Spectrum Access}
\acro{DQL}{Deep Q Learning}
\acro{DRL}{Deep Reinforcement Learning}
\acro{DQN}{Deep Q-Network}
\acro{DDQN}{double deep Q network}
\acro{DDPG}{Deep Deterministic Policy Gradient}
\acro{FBMC}{Filterbank Multicarrier}
\acro{FEC}{Forward Error Correction}
\acro{FFR}{Fractional Frequency Reuse}
\acro{FL}{Federated Learning}
\acro{FSO}{Free Space Optics}
\acro{FANET}{Flying ad-hoc network}
\acro{GA}{Genetic Algorithms}
\acro{GAN}{Generative Adversarial Networks}
\acro{GMMs}{Gaussian mixture models}
\acro{HAP}{High Altitude Platform}
\acro{HL}{Higher Layer}
\acro{HARQ}{Hybrid-Automatic Repeat Request}
\acro{HCA}{Hierarchical Cluster Analysis}
\acro{HO}{Handover}
\acro{KNN}{k-nearest neighbors} 
\acro{IoT}{Internet of Things}
\acro{ISAC}{Integrated Sensing and Communication}
\acro{LAN}{Local Area Network}
\acro{LAP}{Low Altitude Platform}
\acro{LL}{Lower Layer}
\acro{LoS}{Line of Sight}
\acro{LTE}{Long Term Evolution}
\acro{LTE-A}{Long Term Evolution Advanced}
\acro{MAC}{Medium Access Control}
\acro{MAP}{Medium Altitude Platform}
\acro{MDP}{Markov Decision Process}
\acro{ML}{Machine Learning}
\acro{MME}{Mobility Management Entity}
\acro{mmWave}{millimeter Wave}
\acro{MIMO}{Multiple Input Multiple Output}
\acro{NFP}{Network Flying Platform}
\acro{NFPs}{Network Flying Platforms}
\acro{NLoS}{Non-Line of Sight}
\acro{RU}{Radio Unit}
\acro{OFDM}{Orthogonal Frequency Division Multiplexing}
\acro{OSA}{Opportunistic Spectrum Access}
\acro{O-RAN}{Open Radio Access Network}
\acro{C-RAN}{cloud radio access network}
\acro{OMC}{O-RAN Management and Control}
\acro{PAM}{Pulse Amplitude Modulation}
\acro{PAPR}{Peak-to-Average Power Ratio}
\acro{PGW}{Packet Gateway}
\acro{PHY}{physical layer}
\acro{PSO}{Particle Swarm Optimization}
\acro{PU}{Primary User}
\acro{QAM}{Quadrature Amplitude Modulation}
\acro{QoE}{Quality of Experience}
\acro{QoS}{Quality of Service}
\acro{QPSK}{Quadrature Phase Shift Keying}
\acro{RF}{Radio Frequency}
\acro{RIS}{Reconfigurable Intelligent Surface}
\acro{RL}{Reinforcement Learning}
\acro{DRL}{Deep Reinforcement Learning}
\acro{RMSE}{Root Mean Squared Error}
\acro{RN}{Remote Node}
\acro{RRH}{Remote Radio Head}
\acro{RRC}{Radio Resource Control}
\acro{RRU}{Remote Radio Unit}
\acro{RSS}{Received Signal Strength}
\acro{SAR}{synthetic-aperture radar}
\acro{SU}{Secondary User}
\acro{SCBS}{Small Cell Base Station}
\acro{SDN}{Software Defined Network}
\acro{SNR}{Signal-to-Noise Ratio}
\acro{SON}{Self-organising Network}
\acro{SVM}{Support Vector Machine}
\acro{TDD}{Time Division Duplex}
\acro{TD-LTE}{Time Division LTE}
\acro{TDM}{Time Division Multiplexing}
\acro{TDMA}{Time Division Multiple Access}
\acro{TL}{Transfer Learning}
\acro{UE}{User Equipment}
\acro{ULA}{Uniform Linear Array}
\acro{UAV}{Unmanned Aerial Vehicle}
\acro{USRP}{Universal Software Radio Platform}
\acro{XAI}{Explainable AI}
\acro{HCP}{heterogeneous computing platform}
 \acro{IoT}{Internet of Things}
\end{acronym}

\section*{Acknowledgment}
{\footnotesize This work was supported by Engineering and Physical Sciences Research Council United Kingdom (EPSRC), Impact
Acceleration Accounts (IAA), Green Secure and Privacy Aware Wireless Networks for Sustainable Future Connected and Autonomous
Systems, under Grant EP/X525856/1.}
\bibliographystyle{IEEEtran}
\bibliography{references.bib}

\end{document}